\documentclass[aps,prd,showpacs,showkeys,amsmath,amssymb]{revtex4}
\usepackage{graphicx}
\usepackage{bm}
\begin{document}

\title{The $J^P=1^+$ $ud\bar s\bar s$ Tetraquark}
\author{Ying Cui, Xiao-Lin Chen}
\affiliation{Department of Physics, Peking University, Beijing
100871, China}
\author{Wei-Zhen Deng}
\email{dwz@th.phy.pku.edu.cn}\affiliation{Department of Physics,
Peking University, Beijing 100871, China}
\author{Shi-Lin Zhu}
\email{zhusl@th.phy.pku.edu.cn}
\affiliation{Department of
Physics, Peking University, Beijing 100871, China}
\date{\today}
\begin{abstract}
Using the color-magnetic interaction Hamiltonian with SU(3) flavor
symmetry breaking, we perform a schematic study of the masses of
the $J^P=1^+$ tetraquarks in the anti-decuplet representation.
After diagonalizing the mass matrix, we find the $ud\bar s\bar s$
tetraquark lies around 1347 MeV. It decays into $K^+K^0\pi^0,
K^+K^+\pi^-, K^0K^0\pi^+$ via P-wave. The dual suppression from
the not-so-big three-body phase space and P-wave decay barrier may
render this exotic state rather narrow. Future experimental
exclusion of this state will cast doubt on the validity of
applying the simple color-magnetic Hamiltonian to the multiquark
system.
\end{abstract}

\maketitle

\pagenumbering{arabic}

\section{Introduction}
\label{sec1}

Since LEPS collaboration reported the possible existence of the
$\Theta^+$ pentaquark \cite{leps}, there have appeared hundreds of
theoretical papers on this extremely narrow resonance above
threshold. During the past year the experimental evidence of its
existence has weakened \cite{ijmpa,hicks}. Theorists have
speculated various schemes to accommodate both its low mass and
narrow width, among which possible strong correlations between
quarks are quite attractive. This kind of correlation manifests
itself in the form of diquarks \cite{jaffe} or tri-quark clusters
\cite{karliner}.

The quantum number of the so-called ``good" diquark is ${\bar
3}_C, {\bar 3}_F, J^P=0^+$. Except its $J^P$, the good diquark is
very similar to the anti-quark in many respects. In fact, there
exists a kind of supersymmetry between the diquark and anti-quark
\cite{zhu-prc,lich}. For example, replacing the two [ud] diquarks
inside $\Theta$ by two anti-quarks, one arrives at a anti-baryon.
Replacing the $\bar s$ by a third diquark one gets a dibaryon if
such a state exists \cite{zhu-prc}. Similarly, replacing one [ud]
diquark by $\bar s$, one could obtain a tetraquark with double
strangeness.

The identification and interpretation of scalar mesons below 1 GeV
have been a difficult issue in hadron spectroscopy for decades.
Especially their underlying structure and decay patterns have been
challenging. Using only the``good" diquark as the building block,
Jaffe gave a fairly good description of scalar mesons below 1 GeV
assuming they are tetraquarks composed of a pair of diquark and
anti-diquark \cite{belle1,belle2,belle3,belle4}. It's important to
note there does not exist a scalar tetraquark with exotic quantum
numbers in Jaffe's``good" diquark scheme.

Karliner and Lipkin introduced the triquark cluster to lower the
$\Theta$ mass \cite{karliner}. In their scheme, $\Theta$ has the
$(ud\bar{s})-[ud]$ type of configuration with P-wave barrier
between $(ud\bar{s})$ and $[ud]$. Recently Karliner and Lipkin
argued that the $(ud\bar s)$ triquark cluster and anti-strange can
form a narrow resonance with double strangeness \cite{lipkin}. Now
no P-wave is introduced. This state has $I=0, J^P=0^+$. Isospin
symmetry and parity conservation forbids it to decay into either
$K^+K^0$ or $KK\pi$. Hence the only allowed decay mode is
$KK\pi\pi$ if it lies above the four-body threshold. Due to strong
phase space suppression, the width of this tetraquark should be
small according to Ref. \cite{lipkin}. Burns, Close and Dudek
pointed out there would exist a $J^P=1^-$ $ud\bar s\bar s$
tetraquark around 1.6 GeV if the $J^P={1\over 2}^+$ $\Theta^+$
exists \cite{dudek}.

Clearly the existence of such a state is strongly correlated with
the $\Theta$ pentaquark. It will be very interesting to confirm
its existence both theoretically and experimentally. A fully
dynamical calculation of the tetraquark spectrum based on the
first principle is still very demanding, although there have
appeared some preliminary results on tetraquarks from the lattice
QCD approach \cite{suga}.

Recently, in the framework of the flux-tube quark model
Kanada-En'yo, Morimatsu, and Nishikawa argued that the scalar
tetraquark with double strangeness does not exist \cite{belle7}.
Instead they pointed out that the $J^{P}=1^{+}$
$ud\overline{s}\overline{s}$ is stable and low-lying. Its mass is
around 1.4 GeV and decays into $K^\ast\pi$ via S-wave with its
width around 20-50 MeV \cite{belle7}.

On the other hand, there have accumulated good evidence of the
$J^P=1^{-+}$ exotic mesons which can not be $q\bar q$ mesons.
There are two candidates $\Pi_1(1400), \Pi_1(1600)$ \cite{pdg}.
Lattice QCD simulation, QCD sum rule approach and the flux tube
model predict the lowest $1^{-+}$ hybrid meson around $1.9$ GeV.
Their masses are too low as a hybrid meson. The assignment of
$\Pi_1(1400), \Pi_1(1600)$ as tetraquarks with one orbital
excitation is quite attractive now. P-wave tetraquarks were
studied extensively in Ref. \cite{chao1}.

In this work we perform a schematic study of the mass splitting of
the $J^P=1^+$ tetraquarks in the anti-decuplet representation, of
which $ud\bar s \bar s$ is a member. In Section \ref{sec2} we
present the color-magnetic interaction Hamiltonian. Then we
construct the wave functions of $J^P=1^+$ tetraquarks in Section
\ref{sec3}. The formalism of calculating SU(3) flavor symmetry
breaking corrections to the color-magnetic interaction energy is
presented in Section \ref{sec4}. Section \ref{sec5} discusses the
extraction of the parameter and numerical analysis. The last
section is a short summary.

\section{Color-magnetic interaction for the multiquark system}
\label{sec2}

The constituent quark model (CQM) is quite successful in the
description of the meson and baryon spectrum. Within CQM the
color-magnetic interaction arising from one-gluon exchange is
responsible for the mass splitting between the octet and decuplet
baryons as first pointed out by De Rujula, Georgi and Glashow
\cite{belle8}. The Hamiltonian describing the color-spin hyperfine
interaction of a multiquark system reads \cite{belle2,belle5}:
\begin{equation}\label{cm}
H_{CM}=-\sum\limits_{i>j} v_{ij} \overrightarrow{\lambda_{i}}
\cdot\overrightarrow{\lambda_{j}}\overrightarrow{\sigma_{i}}
\cdot\overrightarrow{\sigma_{j}}
\end{equation}
where ${\vec \sigma}_i$ is the quark spin operator and ${\vec
\lambda}_i$ the color operator. For the anti-quark, ${\vec
\lambda}_{\bar q}= -{\vec \lambda}^{*}$ and ${\vec \sigma}_{\bar
q}=-{\vec \sigma}^{*}$. The value of the coefficient $v_{ij}$
depends on the multiquark system and specific models. In the
present convention, $v_{ij}$ is positive. For example, $v_{ij}$
takes different values for $q\bar q$, $qqq$ and $q\bar qq\bar q$
systems. In the bag model, $v_{ij}$ depends on the bag radius and
the constituent quark mass $m_{i,j}$. In CQM, $v_{ij}={\bar
v}\frac{m_u^2}{m_i m_j}$ while ${\bar v}$ depends on the
multiquark system.

Under exact $SU(3)_f$ flavor symmetry, $v_{ij}=v$,
\begin{equation}\label{cm1}
H_{CM}=-v\sum\limits_{i>j} \overrightarrow{\lambda_{i}}
\cdot\overrightarrow{\lambda_{j}}\overrightarrow{\sigma_{i}}
\cdot\overrightarrow{\sigma_{j}}
\end{equation}
We use the notation $|D_{6},D_{3c},S,N,D_{3f}\rangle$ to denote a
particular multiquark configuration, where $D_{6}, D_{3c}$ and
$D_{3f}$ are $SU(6)$ color-spin, $SU(3)_{c}$ color, and
$SU(3)_{f}$ flavor representations of the multiquark system
respectively. S is the spin of the system, and N is the total
number of quarks and antiquarks. The $SU(6)_{cs}$ generators are
$\alpha= \sqrt{\frac{2}{3}}\sigma^{k}, \lambda^{a},
\sigma^{k}\lambda^{a}, \text{k=1,2,3}, \text{a=1,2,...,8} $. For
the antiqauark $\alpha_{\bar q}=-\alpha^{*}$. The Casimir
operators of $SU(6)_{cs}$ and $SU(3)_f$ groups are defined as
$C_{6}=\sum\limits_{u=1}^{35}(\sum\limits_{i=1}^{N}\alpha_{i}^{u})^{2}$,
$C_{3}=\sum\limits_{a=1}^{8}(\sum\limits_{i=1}^{N}\lambda_{i}^{a})^2$.
The color-magnetic interaction energy of the multiquark system can
be expressed in terms of the quadratic Casimir operators of
$SU(2)_{s}, SU(3)_{c}$ ,and $SU(6)_{cs}$.
\begin{equation}\label{master}
H_{CM}=\frac{v}{2}[\overline{C}(\mbox{total})-2\overline{C}(Q)-2\overline{C}(\overline{Q})+16N]
\;,
\end{equation}
where $ \overline{C}=C_{6}-C_{3}-\frac{8}{3}S(S+1) $ and
$\overline{C}(\mbox{total})$, $\overline{C}(Q)$ and
$\overline{C}(\overline{Q})$ denote the $\overline{C}$ of the
whole multiquark system, the subsystem of quarks and antiquarks
respectively. Eq. (\ref{master}) was first derived assuming exact
SU(3) flavor symmetry in Ref. \cite{belle5}.

\section{$1^+$ Tetraquarks}
\label{sec3}

We use diquarks to construct $1^+$ teraquark wave function.
Nominally there are four types of diquarks:
$|21,\overline{3}_{c},0,2,\overline{3}_{f}\rangle,
|21,6_{c},1,2,\overline{3}_{f}\rangle,
|15,\overline{3}_{c},1,2,6_{f}\rangle,
|15,6_{c},0,2,6_{f}\rangle$. The spin and color wave functions of
the first two type of diquarks are simultaneously symmetric or
antisymmetric. Using the master formulae Eq. (\ref{master}),
\begin{equation}
V_{CM}(Q)=-\frac{v}{2}[\overline{C}(Q)-16N]
\;,
\end{equation}
it's easy to show that the color-magnetic interaction is attractive
for the first two types of diquarks and repulsive for the latter two
types of diquarks. Numerically we have $V_{CM}=-8v, -\frac{4}{3}v,
\frac{8}{3}v, 4v$ for four types of diquarks respectively. The first
type of diquark is what Jaffe called the ``good" diquark since its CM
interaction is the strongest. Jaffe used only the good diquark as the
building block of scalar tetraquarks.

We want to emphasize that diquarks are not point-like particles.
Instead they are extended objects in space. There also exists
color-magnetic interaction between quarks inside different
diquarks. We can construct six types of $1^+$ tetraquarks. Their
underlying diquark-diquark structures were presented in Ref.
\cite{belle2}. In the exact $SU(6)$ symmetry limit,
\begin{eqnarray} \label{en1}
|35,1_{c},1,4,(1+8)_{f}\rangle &=&
|21,6_{c},1,2,\overline{3}_{f}\rangle\otimes|\overline{21},\overline{6}_{c},1,2,3_{f}\rangle
\;,\\ \label{en11} |35,1_{c},1,4,(1+8+27)_{f}\rangle &=&
|15,\overline{3}_{c},1,2,6_{f}\rangle\otimes|\overline{15},3_{c},1,2,\overline{6}_{f}\rangle
\;,\\
\label{en1-2} |35,1_{c},1,4,(8+\overline{10})_{f}\rangle &=&
\sqrt{\frac{1}{3}}|21,\overline{3}_{c},0,2,\overline{3}_{f}\rangle
\otimes|\overline{15},3_{c},1,2,\overline{6}_{f}\rangle
-\sqrt{\frac{2}{3}}|21,6_{c},1,2,\overline{3}_{f}\rangle
\otimes|\overline{15},\overline{6}_{c},0,2,\overline{6}_{f}\rangle
\;,\\
\label{en1-3} |280,1_{c},1,4,(8+\overline{10})_{f}\rangle &=&
\sqrt{\frac{2}{3}}|21,\overline{3}_{c},0,2,\overline{3}_{f}\rangle
\otimes|\overline{15},3_{c},1,2,\overline{6}_{f}\rangle
+\sqrt{\frac{1}{3}}|21,6_{c},1,2,\overline{3}_{f}\rangle
\otimes|\overline{15},\overline{6}_{c},0,2,\overline{6}_{f}\rangle
\\
\label{en1-0} |35,1_{c},1,4,(8+10)_{f}\rangle &=&
\sqrt{\frac{1}{3}}|\overline{21},3_{c},0,2,3_{f}\rangle
\otimes|15,\overline{3}_{c},1,2,6_{f}\rangle
-\sqrt{\frac{2}{3}}|\overline{21},\overline{6}_{c},1,2,3_{f}\rangle
\otimes|15,6_{c},0,2,6_{f}\rangle
\;,\\
\label{en1-1}
 |280,1_{c},1,4,(8+10)_{f}\rangle &=&
\sqrt{\frac{2}{3}}|\overline{21},3_{c},0,2,3_{f}\rangle
\otimes|15,\overline{3}_{c},1,2,6_{f}\rangle
+\sqrt{\frac{1}{3}}|\overline{21},\overline{6}_{c},1,2,3_{f}\rangle
\otimes|15,6_{c},0,2,6_{f}\rangle\;, \;.
\end{eqnarray}
In the presence of CM interaction $H_{CM}$ in Eq.~(\ref{cm1}) with
exact flavor sysmetry, the first two kinds of tetraquarks in Eqs.
(\ref{en1})-(\ref{en11}) are still mass eigenstates.
\begin{eqnarray}
|1^+(1+8)_{f}\rangle &=&
|35,1_{c},1,4,(1+8)_{f}\rangle \;,\nonumber\\
|1^+(1+8+27)_{f}\rangle &=&
|35,1_{c},1,4,(1+8+27)_{f}\rangle \;.
\end{eqnarray}
CM interaction $H_{CM}$ mixes the two flavor eigenstates in Eq.
(\ref{en1-2})-(\ref{en1-3}) to yield the mass eigenstates.
\begin{eqnarray} \label{en2}
|1^+(8+\overline{10})_{f}\rangle &=&
(\frac{2\sqrt{2}}{3})|35,1_{c},1,4,(8+\overline{10})_{f}\rangle
 +(\frac{1}{3})|280,1_{c},1,4,(8+\bar{10})_{f}\rangle\;, \nonumber\\
|1^+(8+\overline{10})^{\prime}_{f}\rangle &=&
(\frac{1}{3})|35,1_{c},1,4,(8+\overline{10})_{f}\rangle
 -(\frac{2\sqrt{2}}{3})|280,1_{c},1,4,(8+\bar{10})_{f}\rangle\;.
\end{eqnarray}
$H_{CM}$ also mixes the fifth and sixth states in Eqs.
(\ref{en1-0})-(\ref{en1-1}) to yield the mass eigenstates
$|1^+(8+{10})_{f}\rangle$ and $|1^+(8+{10})^{\prime}_{f}\rangle$.

The CM interaction energy $V_{CM}=-16v , 0,
-\frac{40}{3}v,\frac{32}{3}v,-\frac{40}{3}v,\frac{32}{3}v$ for
$|1^+(1+8)_{f}\rangle$, $|1^+(1+8+27)_{f}\rangle$,
$|1^+(8+\overline{10})_{f}\rangle$,
$|1^+(8+\overline{10})^{\prime}_{f}\rangle$,
$|1^+(8+{10})_{f}\rangle$ and $|1^+(8+{10})^{\prime}_{f}\rangle$
respectively, among which three are unbound with $V_{CM}\ge 0$.
Although $|1^+,(1+8)_{f}\rangle$ and the octet part of
$|1^+(8+\overline{10})_{f}\rangle$ are bound, they do not carry
exotic quantum numbers. They lie around $1.1\sim 1.4$ GeV and mix
strongly with conventional $L=1$ $q\bar q$ states such as
$b_1(1235), a_1(1260), f_1(1285)$ etc. No symmetry forbids them
fall apart into two mesons. Hence these states are very broad.
Experimental identification of these broad bumps above background
is nearly impossible. We do not discuss them further in this work.

From now on, we focus on the anti-decuplet part of the
$|1^+(8+\overline{10})_{f}\rangle$. $|1^+(8+{10})_{f}\rangle$ is
its charge conjugate representation. If $1^+$ tetraquarks {\sl
really} exist, they should belong to this category. In fact, the
anti-decuplet contains several $1^+$ tetraquarks which are exotic
in flavor and useful for the experimental search. Their flavor
wave functions are presented in Table \ref{tab2}. As will be shown
below, the $ud\bar s\bar s$ $1^+$ tetraquark is expected to be
rather narrow from symmetry considerations.
\begin{table}
\begin{center}
\begin{tabular}{|c|c|c|c|}
  \hline
$(Y, I, I_{3})$& Quark content&$V_{CM}(v)$& Mass(MeV)\\
  \hline
$(2,0,0)$&$[ud]\bar s\bar s$&-11.4& 1347     \\
  \hline
$(1,\frac{1}{2},-\frac{1}{2})$&
$\sqrt{\frac{1}{3}}([ds]\bar s\bar s)+\sqrt{\frac{2}{3}}([ud]\{\bar u\bar s\})$&-8.4&1351\\
  \hline
$(1,\frac{1}{2}, \frac{1}{2})$&
$\sqrt{\frac{1}{3}}([us]\bar s\bar s)+\sqrt{\frac{2}{3}}([ud]\{\bar d\bar s\})$&-8.4&1351\\
  \hline
$(0,1,-1)$& $\sqrt{\frac{1}{3}}([ud]\bar
u\overline{u})+\sqrt{\frac{2}{3}}([ds]\{\bar{u}\bar{s}\})$&-10.1&
1256\\
  \hline
$(0,1,0)$& $\sqrt{\frac{1}{3}}([ud]\{\bar{u}\bar{d}\})
+\sqrt{\frac{2}{3}}\sqrt{\frac{1}{2}}([ds]\{\bar{d}\bar{s}\}+[us]\{\bar{u}\bar{s}\})$&-10.1&
1256\\
  \hline
$(0,1,1)$&
$\sqrt{\frac{1}{3}}([ud]\bar{d}\bar{d})+\sqrt{\frac{2}{3}}([us]{\bar{d}\bar{s}})$&-10.1&
1256 \\
  \hline
$(-1,\frac{3}{2},-\frac{3}{2})$&$[ds]\bar{u}\bar{u}$ &-10.1&
1197  \\
  \hline
$(-1,\frac{3}{2},-\frac{1}{2})$&
$\sqrt{\frac{1}{3}}([us]\bar{u}\bar{u})+\sqrt{\frac{2}{3}}([ds]\{\bar{u}\bar{d}\})$
&-10.1&
1197 \\
  \hline
$(-1,\frac{3}{2},\frac{1}{2})$&
$\sqrt{\frac{1}{3}}([ds]\bar{d}\bar{d})+\sqrt{\frac{2}{3}}([us]\{\bar{u}\bar{d}\})$&-10.1&
1197 \\
  \hline
$(-1,\frac{3}{2},\frac{3}{2})$&$ [us]\bar{d}\bar{d}$&-10.1&
1197 \\
  \hline
\end{tabular}
\end{center}
\caption{The flavor wave function and mass of the anti-decuplet
tetraquarks. The CM energy is in unit of v.}\label{tab2}
\end{table}

\section{$1^+$ Tetraquark Masses}
\label{sec4}

For the tetraquark, the Hamiltonian reads
\begin{equation}\label{p1}
H=\sum m(q_i)+H_{CM}
\end{equation}
where $m(q_i)$ is the mass of $i$-th constituent quark. The
$SU(3)_{f}$ flavor symmetry is badly broken since $m_s>m_u$. Besides
the constituent quark mass difference, we need consider the symmetry
breaking corrections to the CM interaction energy in eq.~(\ref{cm}).
The explicit expression of its matrix element between two states
$|i\rangle$ and $|j\rangle$, $V_{CM}=\langle i \mid H_{CM} \mid
j\rangle$ with given flavor context $q_1q_2\bar{q}_3\bar{q}_4$ ($q_i=u,d,s$),
reads
\begin{equation}\label{a0}
 V_{CM}(q_1q_2\bar q_3\bar q_4)=V_{12}(q_{1}q_{2})+V_{13}(q_{1}\bar{q}_{3})
 +V_{14}(q_{1}\bar{q}_{4})+V_{23}(q_{2}\bar{q}_{3})
 +V_{24}(q_{2}\bar{q}_{4})+V_{34}(\bar{q}_{3}\bar{q}_{4})
 \;.
\end{equation}
In the exact flavor symmetry limit,
\begin{equation}\label{HNOCOR}
V_{CM}=V_{12}+V_{13}+V_{14}+V_{23}
+V_{24}+V_{34} \; ,
\end{equation}
which is independent of the flavor context.  With $SU(3)_{f}$ symmetry
breaking corrections $v_{ij}=v\dfrac{m_u^2}{m_im_j}$,
\begin{equation}\label{HCOR}
V_{CM}(q_1q_bar q_bar q_4)=\zeta(q_1)\zeta(q_2)V_{12}
+\zeta(q_1)\zeta(q_3) V_{13}+\zeta(q_1)\zeta(q_4) V_{14}
+\zeta(q_2)\zeta(q_3) V_{23}+\zeta(q_2)\zeta(q_4) V_{24}
+\zeta(q_3)\zeta(q_4) V_{34}
\end{equation}
where $\zeta(q)=\frac{m_{u}}{m_{q}}$.
$|1^+(8+\overline{10})_{f}\rangle$ and
$|1^+(8+\overline{10})^{\prime}_{f}\rangle$ are mass eigenstates
of $H_{CM}$ in the exact SU(3) flavor symmetry in Eq.
(\ref{HNOCOR}). Clearly they are not eigenstates of the CM
interaction with SU(3) flavor symmetry breaking in Eq.
(\ref{HCOR}). They will mix each other and also with the
$|1^+(1+8)_f\rangle$, $|1^+(1+8+27)_f\rangle$, and their conjugate
representation.

In order to get the physical mass eigenstates, we need calculate
every individual term in Eq.  (\ref{HCOR}) and diagonalize the new
$6\times 6$ mass matrix.  For the purpose of calculating $V(q \bar
q)$ terms, we need do some recouplings \cite{belle9, belle10}.
\begin{eqnarray}
\{|q_{1}q_{2}D_{6}(Q)D_{3}(Q)S(Q); \bar q_{3}\bar q_{4}D_{6}(\bar
Q)D_{3}(\bar Q)S(\bar Q)\rangle\}_{(D_{3},S)}\nonumber\\
=\sum R(D_{3}(Q)D_{3}(\bar Q);D_{3}(13)D_{3}(24);D)R(S(Q)S(\bar
Q);S(13)S(24);S)\nonumber\\
\times\{|q_{1}\bar q_{3}D_{6}(13)D_{3}(13)S(13); q_{2}\bar
q_{4}D_{6}(24)D_{3}(24)S(24)\rangle\}_{(D_{3},S)}\;,
\end{eqnarray}
where the recoupling coeffiecients are
\begin{eqnarray}
&&R(S(Q)S(\bar Q);S(13)S(24);S)\nonumber\\
&=&\sqrt{(2S(Q)+1)(2S(\bar
Q)+1)(2S(13)+1)(2S(24)+1)}\bordermatrix{&&&\cr
&\frac{1}{2}&\frac{1}{2}&S(Q)\cr &\frac{1}{2}&\frac{1}{2}&S(\bar
Q) \cr &S(13)&S(24)&S \cr}\;,\\
&&R((\lambda_{Q}\mu_{Q})(\lambda_{\bar Q}\mu_{\bar
Q});(\lambda_{13}\mu_{13})(\lambda_{24}\mu_{24}))\nonumber\\
&=&(-1)^{\lambda_{Q}+\mu_{Q}+\lambda_{13}+\mu_{13}}U((10)(10)(10)(01);
(\lambda_{Q}\mu_{Q})(\lambda_{13}\mu_{13}))\;.
\end{eqnarray}
With $SU(3)$ Racah coefficients
$U((10)(10)(10)(01);(20)(00))=\sqrt{2/3}$,
$U((10)(10)(10)(01);(20)(11))=\sqrt{1/3}$,
$U((10)(10)(10)(01);(01)(00))=-\sqrt{1/3}$,
$U((10)(10)(10)(01);(01)(11))=\sqrt{2/3}$, we have:
\begin{eqnarray}
&&\{|q_{1}q_{2} 21,\bar{3},S=0; \bar q_{3} \bar q_{4} 15,3,S=1
 \rangle\}_{(1,1)}\nonumber\\
&=&\frac{\sqrt{3}}{6}|q_1\bar q_{3} 1,1,0; q_2 \bar q_{4} 35,1,1
 \rangle
 -\frac{\sqrt{6}}{6}|q_1\bar q_{3} 35,8,0; q_2
\bar q_{4} 35,8,1
 \rangle
-\frac{\sqrt{3}}{6}|q_1\bar q_{3} 35,1,1;  q_2 \bar q_{4} 1,1,0
 \rangle\nonumber\\
&+&\frac{\sqrt{6}}{6}|q_1\bar q_{3} 35,8,1; q_2 \bar q_{4} 35,8,0
 \rangle
 -\frac{\sqrt{3}}{3}|q_1\bar q_{3} 35,8,1; q_2
\bar q_{4} 35,8,1
 \rangle
+ \frac{\sqrt{6}}{6}|q_1\bar q_{3} 35,1,1; q_2 \bar q_{4} 35,1,1
 \rangle\;, \\
&&\{|q_{1}q_{2} 21,6,1; \bar q_{3} \bar q_{4} 15,\bar 6,0
 \rangle\}_{(1,1)}\nonumber\\
 &=&-\frac{\sqrt{6}}{6}|q_1\bar q_{3}
1, 1,0;  q_2 \bar q_{4} 35,1,1
 \rangle
 -\frac{\sqrt{3}}{6}|q_1\bar q_{3} 35,8,0; q_2
\bar q_{4} 35,8,1
 \rangle
+\frac{\sqrt{6}}{6}|q_1\bar q_{3} 35,1,1;  q_2 \bar q_{4} 1,1,0
 \rangle\nonumber\\
 &+&\frac{\sqrt{3}}{6}|q_1\bar q_{3} 35,8,1; q_2
\bar q_{4} 35,8,0
 \rangle
+\frac{\sqrt{6}}{6}|q_1\bar q_{3} 35,8,1; q_2 \bar q_{4} 35,8,1
 \rangle
 +\frac{\sqrt{3}}{3}|q_1\bar q_{3} 35,1,1; q_2
\bar q_{4} 35,1,1
 \rangle\;.
\end{eqnarray}
Then we can rewrite the $SU(6)$ flavor eigenstates in terms of two
pairs of $q_1\bar q_3 \otimes q_2\bar q_4$:
\begin{eqnarray} \label{en3}
|35,1_c,1,(8+\bar{10})_{f}
 \rangle
 &=&\frac{1}{2}|q_1\bar q_{3}
1,1,0;  q_2 \bar q_{4} 35,1,1
 \rangle
 -\frac{1}{2}|q_1\bar q_{3}
35,1,1;  q_2 \bar q_{4} 1,1,0
 \rangle\nonumber\\
&-&\frac{2}{3}|q_1\bar q_{3} 35,8,1; q_2 \bar q_{4} 35,8,1
 \rangle
 -\frac{\sqrt{2}}{6}|q_1\bar q_{3} 35,1,1; q_2
\bar q_{4} 35,1,1
 \rangle\;,\\
\label{en4}
|280, 1_c,1,(8+\bar{10})_{f}
 \rangle
 &=&-\frac{1}{2}|q_1\bar q_{3} 35,8,0; q_2
\bar q_{4} 35,8,1
 \rangle
 +\frac{1}{2}|q_1\bar q_{3} 35,8,1; q_2
\bar q_{4} 35,8,0
 \rangle\nonumber\\
&-&\frac{\sqrt{2}}{6}|q_1\bar q_{3} 35,8,1; q_2 \bar q_{4} 35,8,1
 \rangle
 +\frac{2}{3}|q_1\bar q_{3} 35,1,1; q_2
\bar q_{4} 35,1,1
 \rangle\;.
\end{eqnarray}
With the $SU_c(3)$ and $SU_s(2)$ symmetries, the recoupling to
$q_1\bar q_4 \otimes q_2\bar q_3$ can easily obtained.  With the help
of the above formulae, we can calculate the CM energy of any
tetraquark states. The results are collected in Table \ref{tab2} in
unit of $v$.

\section{The Extraction of $v$}\label{sec5}

In order to get the numerical values of $1^+$ tetraquark mass, we
need determine the constituent quark mass and the value of the
parameter $v$ in Eq. (\ref{cm}) for the $1^+$ tetraquark system.
With the experimental values of $\pi, \rho, K, K^\ast$ mesons as
inputs, we use Eq. (\ref{HCOR}) to extract the constituent quark
mass consistently.
\begin{eqnarray}
\begin{cases}
M(\pi)=2m_u-16v' & \\
M(K)=m_u+m_s-16\zeta v'&\\
M(\rho)=2m_u+\frac{16}{3}v'&\\
M(K^{*})=m_u+m_s+\frac{16}{3}\zeta v'&\\
\end{cases}
\Rightarrow\begin{cases} m_u = 308  \mbox{MeV} & \nonumber\\
  m_s = 486 \mbox{MeV} & \nonumber\\
v^{'}= 29.9 \mbox{MeV} & \nonumber\\\zeta=\frac{m_{u}}{m_{s}}
=0.63 &
\end{cases}\nonumber
 \end{eqnarray}
where $v'$ is the CM interaction parameter in Eq. (\ref{cm}) for
the $q\bar{q}$ meson system without the orbital and radial
excitation.

In order to extract $v$ for the tetraquark system, we follow
Jaffe's assumption that scalar mesons below 1 GeV are mainly
composed of a pair of``good" diquark and anti-diquark
\cite{belle1,belle2,belle3,belle4}. We are not arguing this is the
only interpretation. For example, $f_0/a_0(980)$ could be the $KK$
molecule states as suggested by Weinstein and Isgur \cite{belle6}.
They could also be conventional $q\bar q$ states or mixture of
$q\bar q$ and tetraquarks. We use this {\sl working} assumption to
constrain the $1^+$ tetraquark mass only.

Under this working assumption, there exists a scalar tetraquark
nonet \cite{belle1}. Moreover, strong mixing between the SU(3)
singlet scalar tetraquark $\sigma^\prime$ and the octet member
$f_0^\prime$ will split the spectrum, leading to the physical
$f_0(980)$ (${ us \bar u \bar s + ds \bar d \bar s\over
\sqrt{2}}$) and $\sigma$ ($ud \bar u \bar d$). This mixing
mechanism violates OZI rule and can not be described by the
color-magnetic interaction Hamiltonian. Therefore we avoid
$f_0(980)$ and $\sigma$ mesons. Instead we use $a_0(980)$ meson
mass to extract $v$. The quark content of $a_{0}(980)$ with
positive charge is $us\bar d \bar s $. Using similar techniques,
we derive its CM interaction eigenstate.
\begin{eqnarray}
|0^+(1+8)_f\rangle_{(Y=0,I=1)}=0.974|1,1_{c},0,4,(1+8)_{f}\rangle
 -0.225|405,1_{c},0,4,(1+8)_{f}\rangle +\cdots
\end{eqnarray}
where the ellipse denotes the other four components from other
representations with tiny coefficients and
\begin{eqnarray}
|1,1_{c},0,4,(1+8)_{f}\rangle&=&\sqrt{\frac{6}{7}}|21,6_{c},1,2,\overline{3}_{f}\rangle\otimes|\overline{21},\overline{6}_{c},1,2,3_{f}\rangle
+\sqrt{\frac{1}{7}}|21,\overline{3}_{c},0,2,\overline{3}_{f}\rangle\otimes|\overline{21},3_{c},0,2,3_{f}\rangle\nonumber\\
|405,1_{c},0,4,(1+8)_{f}\rangle&=&\sqrt{\frac{1}{7}}|21,6_{c},1,2,\overline{3}_{f}\rangle\otimes|\overline{21},\overline{6}_{c},1,2,3_{f}\rangle
-\sqrt{\frac{6}{7}}|21,\overline{3}_{c},0,2,\overline{3}_{f}\rangle\otimes|\overline{21},3_{c},0,2,3_{f}\rangle
\;.
\end{eqnarray}
Similarly the $SU(6)$ flavor eigenstates can be expressed in terms
of two pairs of $q\bar q$.
\begin{eqnarray}
|1,1_{c},0,4,(1+8)_{f}\rangle
 &=&\frac{\sqrt{21}}{6}|q_1\bar q_{3}
1,1,0;  q_2 \bar q_{4} 1,1,0
 \rangle
 -\frac{\sqrt{7}}{14}|q_1\bar q_{3}
35,1,1;  q_2 \bar q_{4} 35,1,1
 \rangle\nonumber\\
&+&\frac{\sqrt{42}}{21}|q_1\bar q_{3} 35,8,0; q_2 \bar q_{4}
35,8,0\rangle
 -\frac{\sqrt{14}}{7}|q_1\bar q_{3} 35,8,1; q_2
\bar q_{4} 35,8,1
 \rangle\;,\\
|405,1_{c},0,4,(1+8)_{f}\rangle
 &=&-\frac{2\sqrt{42}}{21}|q_1\bar q_{3} 35,1,1; q_2
\bar q_{4} 35,1,1
 \rangle
 +\frac{3\sqrt{7}}{14}|q_1\bar q_{3} 35,8,0; q_2
\bar q_{4} 35,8,0\rangle \nonumber\\
&&  +\frac{5\sqrt{21}}{42}|q_1\bar q_{3} 35,8,1; q_2 \bar q_{4}
35,8,1  \rangle  \;.
 \end{eqnarray}
Its CM interaction energy and mass are
\begin{eqnarray}\nonumber
V_{CM}(a_{0}(980))&=&-28.8v \; ,\\ \nonumber M(a_{0}(980))&=&2m_s
+2 m_u-28.8v \; .
\end{eqnarray}
Using the experimental value of $a_0(980)$ mass as input, we get
\begin{equation}
v \approx21.1 \mbox{MeV} \;.
\end{equation}
The $V_{CM}$ of $\kappa$, $f_0(980)$, and $\sigma$ meson are
$-35.4v$, $-28.8v$ and $-43.4v$ respectively. Thus we can predict
the masses of $\kappa$, $f_0(980)$, and $\sigma$ mesons: $
m_\kappa = 663 \mbox{MeV}$, $ m_{f_{0}(980)} = 980 \mbox{MeV}$,
$m_{\sigma} = 316 \mbox{MeV}$. As pointed out by Jaffe decades ago
\cite{belle1,belle2,belle3,belle4}, strong correlations between
light quarks lead to low-lying scalar tetraquark mesons. Mixing
between flavor eigenstates within different representations
further lowers $\sigma$ meson mass. Since it's above $\pi\pi$
threshold, it falls apart via S-wave easily and becomes very broad
and buried by background. Moreover, one should be cautious that
other complicated mechanisms will alter the sigma meson mass from
the color-magnetic interaction Hamiltonian significantly.

Note the same $v'$ is assumed for $L^P=0^{-}$ and $1^{-}$ $q\bar q$
mesons within this model. In our case neither $0^{+}$ nor $1^{+}$
tetraquarks have orbital excitation. Hence the parameter $v$
should be the same for both $0^{+}$ and $1^{+}$ tetraquark
systems. Now we can estimate the $1^{+}$ tetraquark mass. For
example, the mass of the $A_{1, 333}$ member of the anti-decuplet
($ud \bar s \bar s$) reads
\begin{eqnarray}
M(A_{1, 333})&=& \sum m_q + V_{CM}(A_{1, 333})=m(u)+m(d)+2m(s)+V_{CM}(A_{1, 333}) \nonumber\\
&=& 308\times2 + 486\times2+(-11.4\times21.1)\approx1347\mbox{MeV}
\; .
\end{eqnarray}
The anti-decuplet $1^+$ tetraquark masses are collected in Table
\ref{tab2}.

\section{Discussions}\label{sec6}

In short summary, we have performed a schematic study of the
masses of $1^+$ tetraquarks in the decuplet representation. We
have paid special attention to the SU(3) flavor symmetry breaking
corrections to the color-magnetic interaction energy. The SU(3)
flavor symmetry breaking color-magnetic Hamiltonian mixes
$J^P=1^+$ states in different SU(6) representations. We have
diagonalized the $6\times 6$ mass matrix to obtain the masses of
$1^+$ tetraquarks in the decuplet representation.

Only three flavor-exotic $1^+$ tetraquarks $ud\bar s\bar s$,
$us\bar d\bar d$, $ds\bar u\bar u$ are potentially interesting.
All the other seven states will easily fall apart into two mesons
via S-wave and become completely buried by background. Among the
three flavor exotics, both $us\bar d\bar d$ and $ds\bar u\bar u$
$1^+$ tetraquarks will unfortunately fall apart into $K^\ast \pi$
very easily via S-wave. Hence they are too broad too. It's
impossible to measure these states experimentally.

Now let's focus on $1^+$ $ud\bar s\bar s$ tetraquark. Let's call
it ${\cal T}^+$. Its quantum numbers are: Y=+2, Q=+1, I=0. With a
mass of 1347 MeV, $ud\bar s\bar s$ can not decay into $K^\ast K$
final states since $m_{K^\ast}+m_K=1386$ MeV. Parity and angular
momentum conservation forbid it decay into $K^+K^0,
K^+K^+\pi^-\pi^0, K^+K^0\pi^0\pi^0, K^0K^0\pi^+\pi^0$. Its decay
modes, which are allowed by both kinematics and symmetry, are
$K^+K^0\pi^0, K^+K^+\pi^-, K^0K^0\pi^+$. These decays occur
through P-wave. The dual suppression from the not-so-big
three-body phase space and P-wave decay barrier may render this
exotic state rather narrow.

We strongly call for experimentalists to search for this
interesting state (1) in the double-strangeness-exchange reactions
on the proton or deuteron target: $K^+ d\to p+p + K^- +{\cal
T}^+$; or (2) in the $J/\psi$ or $\Upsilon$ decays: $J/\psi
\left(\Upsilon\right)\to K^-{\bar K}^0 {\cal T}^+$. The existence
of this state is a generic feature of the color-magnetic
interaction model. The future experimental exclusion of this state
will cast doubt on the validity of the application of the simple
color-magnetic Hamiltonian to the multiquark system.

\section{Acknowledgments}

This project was supported by the National Natural Science
Foundation of China under Grants 10375003 and 10421003, Ministry
of Education of China, FANEDD, Key Grant Project of Chinese
Ministry of Education (NO 305001) and SRF for ROCS, SEM. C.Y.
thanks Y. R. Liu and W. Wei for helpful discussions.

\end{document}